\documentclass[twocolumn,showpacs,preprintnumbers,amsmath,amssymb]{revtex4}
\usepackage{tabularx,graphicx}

\usepackage{color}
\usepackage{hyperref}
\hypersetup{
    colorlinks=true,
    linkcolor=blue,
    filecolor=blue,      
    urlcolor=blue,
}

\begin{document}

\newcommand{\beq}{\begin{equation}}
\newcommand{\eeq}{\end{equation}}
\newcommand{\beqn}{\begin{eqnarray}}
\newcommand{\eeqn}{\end{eqnarray}}
\newcommand{\bmath}{\begin{subequations}}
\newcommand{\emath}{\end{subequations}}
\newcommand{\bra}[1]{\langle #1|}
\newcommand{\ket}[1]{|#1\rangle}

\title{On the reversibitity of the Meissner effect and the angular momentum puzzle}
\author{J. E. Hirsch }
\address{Department of Physics, University of California, San Diego,
La Jolla, CA 92093-0319}

\begin{abstract} 
It is generally believed that the laws of  thermodynamics govern superconductivity as
an equilibrium state of matter. Here we point out that
within the conventional   BCS-London description of   the normal-superconductor
transition   in the presence of a magnetic field, the transition 
cannot  be reversible, in contradiction with the thermodynamic description and with
experiments.  This indicates that the conventional theory of superconductivity is internally inconsistent. 
We argue  that to describe a reversible transition it is necessary to assume that charge transfer occurs
across the normal-superconductor phase boundary, as proposed in the theory of hole superconductiviy. 
This provides also a solution to the angular momentum puzzle pointed out in previous work.
Our explanation can only apply if the current carriers in the normal state are holes.
An experimental test of these ideas is proposed.
 \end{abstract}
\maketitle

\section{introduction}

The experimental discovery of the Meissner effect in 1933 \cite{meissner} suggested that the transition
between normal and superconducting states in the presence of a magnetic field is a reversible phase transformation
between well-defined equilibrium states of matter
to which the ordinary laws of equilibrium thermodynamics apply \cite{gc}. For example,  
the Rutgers relation \cite{rutgers} relating the specific heat jump  between normal and superconducting
phases  at the critical temperature to the  
 temperature derivative of the thermodynamic critical field follows from
this description. In fact, the Rutgers relation had been found experimentally and interpreted theoretically using thermodynamics
\cite{gorter} before
the discovery of the Meissner effect, in a sense anticipating it.
Subsequent extensive experimental tests \cite{keesom,keesom2} confirmed that in the ideal situation the normal-superconductor transition occurs
without entropy production within experimental accuracy, i.e. is 
reversible, and this has been generally believed ever since.

In this paper we point out that within the conventional London-BCS theory of superconductivity \cite{tinkham} the transition between normal
and superconducting states in the presence of a field  $cannot$ be   reversible but instead is necessarily 
associated with entropy production.
If so, this would render the usual thermodynamic description invalid, and
  indicate that the experiments consistent with reversibility were flawed \cite{keesom,keesom2,map}. However, we argue instead  that the conventional London-BCS description of the transition is flawed,
  and that the transition $is$ reversible and thermodynamics applies because of some physics 
  that is absent in London-BCS theory but occurs in nature during the normal-superconductor
  transition:   $charge$ $transfer$ in
  direction perpendicular to the normal-superconductor phase boundary.

  In recent work we have argued that  charge transfer in
  direction perpendicular to the normal-superconductor phase boundary is necessary to explain the dynamics
  of the Meissner effect \cite{dyn1,dyn2}. In this paper we show that this charge transfer is necessary to render the
  transition between normal and superconducting states reversible, and that the transition would be irreversible in the
  absence of this charge transfer, in contradiction with experiment. In addition, we show that this charge transfer resolves the 
  angular momentum puzzle associated with the Meissner effect that we pointed out in previous work \cite{ang1,ang2}.
  Finally, we discuss an experimental test of these ideas.

          \begin{figure}
 \resizebox{8.5cm}{!}{\includegraphics[width=6cm]{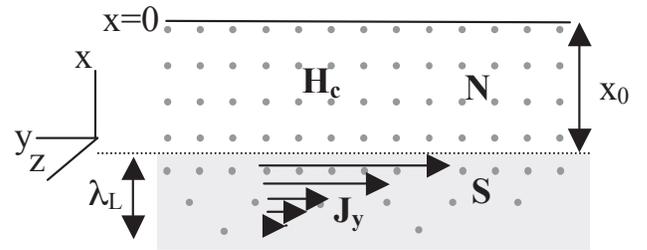}}
 \caption {  Normal (N) and superconducting (S)  phases in equilibrium. The N-S phase boundary is shown by the dotted line. The magnetic field 
 $H_c$ (indicated by grey dots) points out of the paper. The current flowing in the (-y) direction  in the superconducting region within a London penetration depth ($\lambda_L$) of the phase boundary 
 screens the
 magnetic field so that it is $0$ beyond  $\lambda_L$ of the phase boundary. No current flows in the normal region. }
 \label{figure1}
 \end{figure}

  \section{Phase equilibrium}
  In a seminal paper \cite{londonh}, H. London analyzed the phase equilibrium between normal and superconducting states in the presence of a
  magnetic field. The situation is shown schematically in Fig. 1.   
  Following the   treatment and notation of ref. \cite{dyn1},   in the superconducting phase ($x<x_0$) a current flows along the $y$ direction parallel to the phase boundary located at $x=x_0$,  given by
  \beq
  J_y(x)=-\frac{c}{4\pi \lambda_L}H_ce^{(x-x_0)/\lambda_L}
  \eeq
 and correspondingly the magnetic field in this region $\vec{H}(x)=H(x)\hat{z}$ is given by
  \beq
  H(x)=H_ce^{(x-x_0)/\lambda_L}
  \eeq
  so as to satisfy the London and Ampere equations
  \bmath
  \beq
\vec{\nabla}\times \vec{J}=-\frac{c}{4\pi\lambda_L^2}\vec{H}==>
\frac{\partial J_y}{\partial x}=-\frac{c}{4\pi \lambda_L^2}H
\eeq\beq
\vec{\nabla}\times \vec{H}=\frac{4\pi}{c}\vec{J}==>
\frac{\partial H}{\partial x}=-\frac{4\pi}{c}J_y .
\eeq
\emath
The system is at temperature $T<T_c$ and  the thermodynamic critical
field at that temperature is $H_c$.
With the current given by $J_y=e n_sv_s$, $n_s$ the number of superconducting carriers  of charge $e$ per unit volume, 
and using the standard relation $1/ \lambda_L^2=4\pi n_s e^2/(mc^2)$ \cite{tinkham} the kinetic energy of the
supercurrent per unit volume at $x=x_0$ is, from Eq. (1)
\beq
\epsilon_k=n_s\frac{1}{2}m v_s^2=\frac{m J_y^2}{2e^2} =\frac{H_c^2}{8\pi} .
\eeq 
 London \cite{londonh} considered a virtual displacement of the phase boundary and derived as equilibrium condition for coexistence of the two phases
    \beq
 \Delta F= \frac{H_c^2}{8\pi} =n_s\frac{1}{2}m v_s^2
  \eeq
  where $\Delta F$ is the difference in free energy per unit volume between the normal and the superconducting phase.
  For the case where the
  phase boundary moves into the superconducting region the situation  is shown schematically in Fig. 2. 
Eq. (5) says that  the kinetic energy of the supercurrent at the phase boundary equals the superconducting condensation energy, and no
energy is lost in irreversible processes.
  In this derivation London neglected the Joule heat $Q$  that would necessarily be generated when the phase boundary
  is displaced and an electric field is induced according to Faraday's law, arguing that if the motion of  the phase boundary   is slow enough it can be neglected and hence the problem could be treated as a
 reversible phase transformation.

However, we argue that London's analysis is flawed because Joule heat is necessarily generated when the
phase boundary moves, no matter
how slowly the process is. Since Joule heat is only generated in the normal region, more energy will be
dissipated when the phase boundary moves into the superconducting region and the normal region is
enlarged than in the reverse process. Thus,   the phase boundary will 
spontaneously move in the direction of enlarging the superconducting region and thermodynamic
equilibrium will not result under the `equilibrium condition' Eq. (5).

                      \begin{figure}
 \resizebox{7.5cm}{!}{\includegraphics[width=6cm]{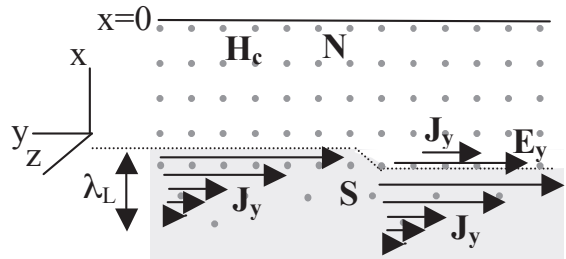}}
 \caption {  The N-S phase boundary (dotted line) is displaced by a small amount in the negative $\hat{x}$ direction,
 enlarging the normal region. 
 This causes an increase   in the magnetic flux in that region and 
 an electric field $E_y$ is necessarily generated, producing a current $J_y=\sigma E_y$ in the region becoming 
 normal ($\sigma=$ normal state conductivity).
 }
 \label{figure1}
 \end{figure}

    \section{Reversibility and entropy production }
    In thermodynamics a sharp distinction is made between quasistatic and reversible processes.  Both processes
    have to happen slowly enough that the system can be considered to be in thermodynamic equilibrium at
    all times, so reversible processes are also quasistatic. However, a quasistatic process will be irreversible if it gives rise to an entropy increase.
    Instead in a reversible process  entropy is not generated at any time.
    
    We argue that the normal-superconductor transition in the presence of a magnetic field cannot be reversible within the conventional
    description.
    As the phase boundary is displaced and the magnetic flux changes, a Faraday field $E_y$ will $necessarily$ be generated
    which will give rise to a current $J_y$ in the normal region, as shown schematically in Fig. 2.   
       Thus, the normal-superconductor transition in the conventional understanding is necessarily irreversible,
    even if it can be argued that the irreversibility is very small if the transition occurs very slowly \cite{londonh}.
    However, the speed at which the transition occurs is determined by the physical conditions
    in the system \cite{pippard,dyn1}. In the next section we will estimate the degree of irreversibility that is
    expected in a typical situation and argue that it is incompatible with the numerous experimental
    reports that appeared to confirm reversibility of the normal-superconductor transition in a
    magnetic field.

        \section{Quantitative estimate of irreversibility}
      Consider the rectangular geometry   shown in Fig. 3. 
      The magnetic field points out of the paper and we assume for simplicity that it
      has magnitude $H_c$ over the entire 
      normal region of depth $x_0$. Assume the phase boundary moves up at a uniform
      rate $\dot{x}_0$, so that  the magnetic field is expelled over a time
      \beq
      t_R=\frac{R}{\dot{x}_0}
      \eeq  
        where R is the initial length of the normal region in the $x$ direction. The phase boundary position
        at time $t$  is given by
        \beq
        x_0(t)= \dot{x}_0t -R
        \eeq 
        According to Faraday's law, an electric field exists in the $y$ direction in the normal region given by
        \beq
        E_y=\frac{H_c}{c}\dot{x}_0
        \eeq
        giving rise to a current $J_y=\sigma E_y$, with $\sigma$ the normal state conductivity. The total Joule heat dissipated per
        unit time is
        \beq
        w(t)=J_yE_y[x_0(t)A]=\sigma E_y^2 [x_0(t)A]
        \eeq
        over a volume $V(t)=x_0(t)A$, with $A$ the cross-sectional area of the sample. Integrating over time we obtain for the total heat dissipated per
        unit volume
        \beq
        W=\frac{1}{RA}\int_0^{t_R} w(t)=\frac{\sigma E_y^2}{2}t_R=\frac{H_c^2}{8\pi}\frac{4\pi \sigma}{c^2}\frac{R^2}{t_R}  .
        \eeq
Using a Drude form for $\sigma$
        \beq
        \sigma=\frac{ne^2\tau}{m_e}
        \eeq
        with $\tau$ the Drude collision time, together with the usual relation for the London penetration depth \cite{tinkham}
        \beq
        \frac{1}{\lambda_L^2}=\frac{4\pi n e^2}{m_e c^2}
        \eeq
        we can write Eq. (10) as
        \beq
        W=\frac{H_c^2}{8\pi} (\frac{R}{\lambda_L})^2 \frac{\tau}{t_R}
        \eeq
        For example, for $R=1cm$, $\lambda_L=500\AA$ 
        \beq
        W=\frac{H_c}{8\pi}(4\times10^{10})\frac{\tau}{t_R}
        \eeq
        so that for a typical   collision time at low temperatures $\tau=10^{-11}s$ 
         if it took $t_R=10s$ for the phase boundary to move $1cm$ the
        Joule heat dissipated is $4\%$ of the condensation energy. 
         In typical experiments the time to reach equilibrium is several seconds, so this simple calculation 
        suggests that the heat dissipated in this irreversible process can be an appreciable fraction
        of the condensation energy. This appears inconsistent with the experimental reports that indicate
        near perfect reversibility in these processes \cite{keesom,keesom2, map}.
        
             \begin{figure}
 \resizebox{8.5cm}{!}{\includegraphics[width=6cm]{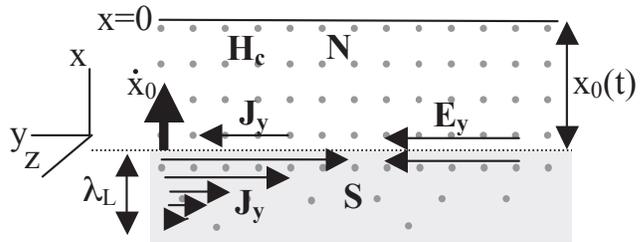}}
 \caption {  The N-S phase boundary (dotted line) is moving up with speed $\dot{x}_0$. The magnetic field (indicated by grey dots) points out of the paper. 
 A Faraday field $E_y$ is induced, which produces a current $J_y=\sigma E_y$ in the normal region flowing parallel to the phase boundary in the $+\hat{y}$ direction. 
  }
 \label{figure1}
 \end{figure} 
        
        We can obtain a more accurate estimate of the energy dissipated using the procedure
        of Refs. \cite{pippard,dyn1}. We assume the system is expelling a magnetic field $H_c(1-p)$ from its
        interior, with $p>0$. As the phase boundary advances, the eddy currents induced raise the
        magnetic field to $H_c$ at the phase boundary, and this limits the speed of growth of the
        superconducting phase. The time evolution of the phase boundary is given by \cite{dyn1}
        \beq
        x_0(t)^2=R^2-\frac{\alpha p c^2}{2\pi \sigma} t
        \eeq
        where the parameter $\alpha$ is determined by the condition
        \beq 
        \alpha\int_0^1dye^{\frac{\alpha p}{2}(y^2-1)}=1 .
        \eeq
        The induced current in the normal region at position x and time t  is given by \cite{dyn1}
        \beq
        J_y(x,t)=-\frac{c}{4\pi}\frac{\alpha p}{x_0}H_ce^{\frac{\alpha p}{2}((x/x_0(t))^2-1)}
        \eeq
        and the total time for the phase boundary to move from $x_0=-R$ to $x_0=0$ is given by
        \beq
        t_R=\frac{2\pi \sigma}{\alpha p c^2}R^2 .
        \eeq
        The energy dissipated per unit volume is
        \beq
        W=\frac{1}{R}\int_0^{t_R}dt\int_{x_0(t)}^0dxJ_y(x,t)E_y(x,t)
        \eeq
        with $E_y=J_y/\sigma$, and a straightforward calculation yields
        \beq
        W=2p\frac{H_c^2}{8\pi} .
        \eeq
         Note that this result is very similar to what was obtained in the earlier calculation, since
         replacing $p$ in Eq. (20) in terms of $t_R$ (Eq. (18) yields
         \beq
         W=\frac{H_c^2}{8\pi}\frac{4\pi \sigma}{\alpha c^2}\frac{R^2}{t_R} 
         \eeq
         the same as Eq. (10) except for the parameter $\alpha$, which approaches
         $1$ for $p\rightarrow 0$. For small $p$, $\alpha=3/(3-p)$ \cite{dyn1}.

         Eq. (20) indicates that if the system is in the normal state at temperature $T<T_c$ in a magnetic field
         slightly larger than $H_c(T)$ and the magnetic field is lowered for example  to $0.95 H_c$,
         in the process of becoming superconducting and expelling the magnetic field,
         an entire $10\%$ of the condensation energy of the superconductor
         will be dissipated as heat. The same result Eq. (20) is obtained for the reverse process
         where the system is initially in the superconducting state in a magnetic field
         smaller than $H_c$ and the magnetic field is increased to $H_c(1+p)$, causing
         the phase boundary to advance into the superconducting phase.
  The calculation  assumes that  the supercurrent is $not$ dissipated
         in Joule  heat when a region goes normal, however Joule heat is generated in the normal
         region because of the Faraday electric field resulting from the changing magnetic field.
         These results show that the transition from the superconducting to the normal state in the presence of a magnetic field, 
         as well as the transition from the normal to the superconducting state in the presence of a magnetic field,
         cannot be reversible since the irreversible heat dissipated is given by Eq. (20) which is non-zero for any value
         of $p$, except $p=0$ where no transition occurs. The irreversibility becomes smaller as the parameter $p$ decreases
         and the time $t_R$ over which the transition occurs Eq. (18) increases, but is always nonzero for any finite $t_R$.
         
         The results discussed above were for a planar geometry where the calculations are
         simplest, however we have found that the
         results are very similar for a cylindrical geometry \cite{dyn1}, and it is to be expected also in other geometries.
         Whether the system becomes superconducting through the process of expansion of a single
         domain as calculated here or through the (more realistic) process of creation of superconducting kernels in many
         locations at random that expand and merge, should not change the results. The essential fact is that
         to expel the magnetic field from the interior of a simply connected superconducting body, the magnetic
         field lines have to move through the entire body to the surface and the energy dissipation will only
         depend on the speed of the process and not on the details of domain growth.

  \section{the puzzle}
  The derivation of the  relation between change in entropy per unit volume and 
  temperature derivative of critical field \cite{gorter}
    \beq
S_n-S_s=-\frac{1}{8\pi}\frac{\partial H_c^2}{\partial T}
  \eeq
  rests on the assumption
  that the heat transfer to render the superconductor normal is given by
  \beq
  \delta Q=TdS
  \eeq
  where $dS$ is the change in entropy. In other words, that the process is reversible. The   Rutgers relation  
  for the specific heat jump at $T_c$
   \beq
  C_s(T_c)-C_n(T_c)=\frac{T_c}{4\pi}(\frac{\partial H_c}{\partial T})_{T=T_c}^2
  \eeq 
  follows from this equation. 
  In addition, at any temperature $T<T_c$ the latent heat involved in the normal-superconductor transition
  for a sample of volume V is
  \beq
  Q=T(S_n-S_s)V=-VT\frac{H_c}{4\pi}\frac{\partial H_c}{\partial T}
  \eeq
  assuming no irreversible increase in entropy takes place during the transition. 
  
  Keesom and coworkers did careful tests of the relation Eq. (25) for both the superconductor-normal \cite{keesom2} and the
  normal-superconductor \cite{keesom} transitions, and found that it holds to great accuracy.   
  As an example, they used an ellipsoidal sample of $Sn$ ($T_c=3.72K$) of dimensions 17.5 cm x 3.5 cm and  measured the latent heat 
  in the S-N transition at temperature $1.239K$.  Assuming for the conductivity of $Sn$ at low temperatures   $\sigma=5\times 10^8 \Omega^{-1} cm^{-1}$ \cite{pippard}
 yields for Eq. (18)
  \beq
  t_R=\frac{\pi}{p} R(cm)^2 s
  \eeq
  as the time it takes for the transition to take place in applied magnetic field $H_c(1+p)$ for sample
  dimension $R$. According to Eq. (20), this should
  result in an irreversible heat dissipation of $W=2pH_c^2/8\pi$, or approximately
  \beq
  W\sim p\frac{T_c}{T} Q
  \eeq
  with $Q$ given by Eq. (25). According to Keeson and van Laer \cite{keesom2}, Eq. (25) was satisfied to
  $0.1\%$ accuracy in their experiment, which implies from Eq. (27) $pT_c/T<0.001$, hence $p<0.00033$. 
  From Eq. (26) and assuming $R=1.75cm$ (half the diameter of the ellipsoid) yields
  \beq
  t_R=16,660 s
  \eeq or $4.6$ hours for the duration of the experiment. Instead, according to Ref. \cite{keesom2}, the 
  experiment took only $696 s$. Conversely, we conclude that if the experiment took $696s$
  the difference in the two sides of Eq. (25) should have been a factor of 24 larger than found by 
  Keesom and van Laer. If the experiment was performed with a higher purity sample of $Sn$ with up to 
  two orders of magnitude smaller resistivity \cite{tinres}, the experiments would have to 
  extend over 460 hours to show reversibility to $0.1\%$ accuracy according to these estimates.

In summary, the experiments showed \cite{keesom,keesom2} that the normal-superconductor transition is reversible to an 
accuracy much larger than expected for the duration of the experiments if eddy currents are generated
in the normal region as the magnetic field changes as predicted by Faraday's law. In other words,
 the transition should be expected to exhibit far more irreversibility than was found in practice. How this
can be explained is discussed in the next sections.

      \section{how to render the normal-superconductor transition reversible}
      Consider again motion of the phase boundary into the normal region at speed $\dot{x}_0$. 
      We cannot negate Faraday's law, so an electric field $E_y=(H_c/c)\dot{x}_0$ will necessarily be generated at the
      boundary and by continuity in the normal region nearby. Is it possible that $no$ current $J_y$ is generated on the normal side of the phase boundary so as to avoid the irreversible Joule heating $J_yE_y$ resulting from it?

          \begin{figure}
 \resizebox{8.5cm}{!}{\includegraphics[width=6cm]{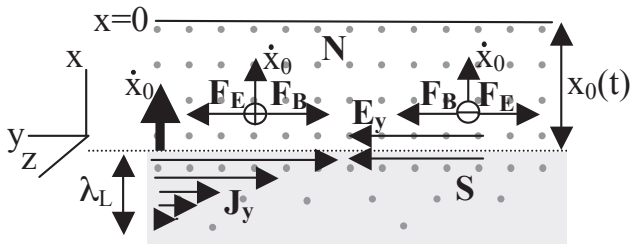}}
 \caption {  The N-S phase boundary (dotted line) is moving up with speed $\dot{x}_0$. The magnetic field (indicated by grey dots) points out of the paper. 
 A Faraday field $E_y$ is induced.  
  If mobile charges (either electrons or holes) are drifting up at speed $\dot{x}_0$  they experience no net force in direction
 parallel to the phase boundary and hence no current $J_y$ is generated.
  }
 \label{figure1}
 \end{figure}

 The answer is yes, provided the normal phase mobile charges are drifting at speed $\dot{x}_0$ in the same direction as the phase boundary.
 If so, the electric and magnetic forces are exactly balanced, both for positive and negative charges (holes or electrons) as shown
 schematically in Fig. 4, since for a charge $q$
 \beq
 \vec{F}_E\equiv qE_y\hat{y}=-\frac{q}{c}\dot{x}_0H_c\hat{x}\times\hat{z}\equiv -\vec{F}_B
 \eeq
 As a consequence, the phase boundary can move into the normal region as the superconducting phase expands $without$ generation
 of Joule heat. The same argument holds for the opposite motion of the phase boundary into the superconducting region as the normal phase
 expands, with the normal charges now moving in opposite
 direction again following the motion of the phase boundary.
 
 The flow of normal charges in direction perpendicular to the phase boundary
 depicted in Fig. 4 is precisely what is expected within the explanation of the Meissner effect provided by the theory
 of hole superconductivity \cite{dyn1,dyn2}, as discussed in the next section.
       
      \section{Orbit expansion and backflow}
      
      We have argued  in previous work that the perfect diamagnetism of superconductors implies that superfluid electrons reside in mesoscopic orbits
      of radius $2\lambda_L$ \cite{ang1,bohr}. The idea that superconducting carriers reside in large orbits was also proposed by
      several researchers in the pre-BCS era \cite{orbits1,orbits2,orbits3}.  The Larmor diamagnetic susceptibility for electrons in orbits
      of radius $k_F^{-1}$ and radius $2\lambda_L$ respectively yields the Landau diamagnetic susceptibility of normal metals and $(-1/4\pi)$, perfect diamagnetism, 
      respectively \cite{ang2}, as appropriate for the normal and superconducting phases.
       This suggests that in the transition to the superconducting state, carriers expand their orbits from microscopic radius $k_F^{-1}$ to radius $2\lambda_L$, and as the magnetic flux through the enlarging orbit increases the carrier acquires an azimuthal velocity 
      generating a magnetic field opposite to the applied one. This  provides a $dynamical$ explanation for the origin of the Meissner current,
      that is not provided within conventional BCS theory \cite{dyn2}.
      In the planar geometry considered here, the orbits in the normal and superconducting state are shown schematically in Fig. 5. Note that 
      the enlarged orbits   in the superconducting region with centers at distance less that $2\lambda_L$ from the boundary enter partially
      into the normal region, up to a distance $2\lambda_L$ from the boundary into the normal region.
      Assuming the orbits correspond to electrons, this implies that negative charge enters from the
      superconducting into the normal region as an orbit next to the phase boundary expands
      
                \begin{figure}
 \resizebox{8.5cm}{!}{\includegraphics[width=6cm]{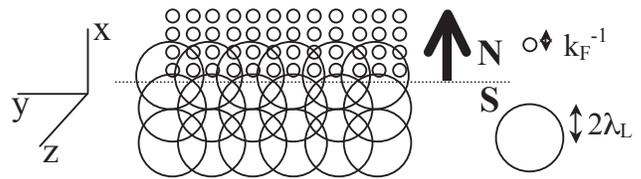}}
 \caption {Electronic orbits in the normal and superconducting states as proposed in the theory of hole superconductivity \cite{ang2}.
 Note that  large orbits centered  in the superconducting side close to the phase boundary enter the normal side up to a distance
 $2\lambda_L$ from the phase boundary.  }
 \label{figure1}
 \end{figure} 
 
                            \begin{figure}
 \resizebox{8.5cm}{!}{\includegraphics[width=6cm]{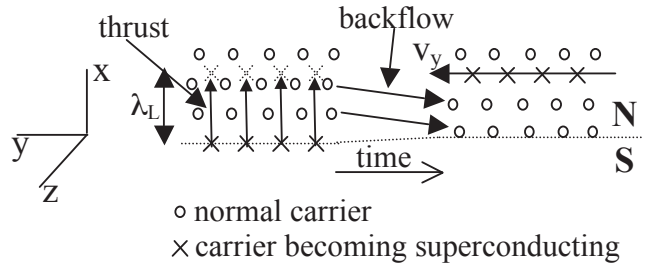}}
 \caption {  As a consequence of the orbit enlargement when carriers become superconducting, negative charge thrusts from the superconducting
 into the normal region and in the process acquires through the Lorentz force a velocity $\vec{v}_y$ in the $+\hat{y}$ direction. In addition,
 this generates a backflow of charge in the normal region. 
  }
 \label{figure1}
 \end{figure}

 The consequence of this orbit enlargement as carriers become superconducting for the process where the phase boundary moves into
 the normal region is shown schematically in Fig. 6 \cite{dyn1}.  As new carriers (electrons) become superconducting and their orbits enlarge, 
 negative charge thrusts into the positive x direction and acquires through the Lorentz force momentum in the $+\hat{y}$ direction,
 resulting in speed $v_y$.  The Lorentz force acting on an electron thrusting forward with speed $v_x$  is
      \beq
           \vec{F}_L=-\frac{e}{c}v_xH_c \hat{y}
           \eeq
        and the speed in the $\hat{y}$ direction that an electron acquires in time $\Delta t$ is
        \beq
        v_y=\int   _0^{\Delta t}\frac{F_L}{m_e} dt=-\frac{e}{c}H_c \int   _0^{\Delta t}v_x dt=-\frac{e}{c}H_c\Delta x
        \eeq
        so that for $\Delta x=\lambda_L$
        \beq 
        v_y=-\frac{e}{m_e c} \lambda_L H_c
        \eeq           
           which is precisely the speed of the carriers in the Meissner current Eq. (1).
           Under the assumption that $v_x >> \dot{x}_0$, the effect of $E_y$ on the forward thrusting electron can be ignored.
           This process then explains what drives the generation of the   Meissner current flowing against the Faraday field $E_y$ as the superconducting phase 
           boundary advances into the normal phase.     
           This physics also explains how in the reverse process, when the normal phase advances
  into the superconducting phase, the Meissner current stops without generating Joule heat, as will be discussed in
  detail in Sect. X.

                                                 \begin{figure}
 \resizebox{8.5cm}{!}{\includegraphics[width=6cm]{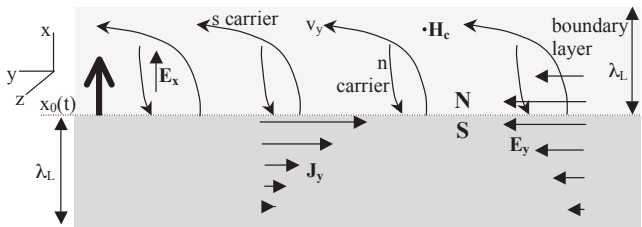}}
 \caption {  As carriers become superconducting (s carriers) they thrust
forward into the normal region over a boundary layer of thickness
$\lambda_L$, and are deflected by the Lorentz force acquiring speed $v_y=-c/(4\pi n_s e \lambda_L)H_c$,
in the $+\hat{y}$ direction assuming the s carriers are electrons. 
 This process creates an electric field $E_x$ in
the $+\hat{x}$  direction that drives normal carrier (n carrier) backflow.  
  }
 \label{figure1}
 \end{figure} 
  
       This motion of negative (superconducting) charge into the normal region will create a charge imbalance and  an electric field $E_x$ will be generated in the normal region within distance $\lambda_L$ from the phase
           boundary pointing in the $+\hat{x}$ direction, that will drive a flow of normal charge  in the $x$ direction, as shown schematically in Fig. 7.
           Fig. 7, reproduced from Ref. \cite{dyn2}, assumes that the normal carriers (n carriers) are negatively charged electrons. This is incorrect, 
           as discussed in the next section.
           In addition, the caption of this figure in Ref. \cite{dyn2} read ``The normal (n) carriers do
not acquire a large $v_y$  in the opposite direction because they scatter
off impurities and transfer their y-momentum to the lattice.'' This is also incorrect, as we discuss in the subsequent section.

  \section{The sign of the normal state charge carriers}
  
  Experiments that measure the gyromagnetic effect \cite{gyro}. the London moment \cite{lm} and the Bernoulli potential \cite{bern} in superconductors establish that the superconducting
  charge carriers are negatively charged. Therefore, as carriers at the boundary become superconducting $negative$  $charge$ is transferred into the 
  normal region within a boundary layer, creating an electric field $E_x$ pointing in the $+\hat{x}$ direction and a normal `backflow' current $J_x$ flowing in the
$+\hat{x}$ direction. This backflow of normal carriers could occur through negative $electrons$ moving
 in the $-\hat{x}$ direction or through positive $holes$ moving in the $+\hat{x}$ direction.
 
 Figure 8 shows the forces acting on the backflowing normal carriers. 
 The speed of the normal carriers in the $x$ direction, $v_x$, has to be $\dot{x}_0$, the speed of motion of the
 phase boundary,  so that no charge
accumulation results. If the normal carriers are electrons, electric 
 ($F_E$) and magnetic ($F_B$) Lorentz forces
 act in the same direction ($-\hat{y}$)  as shown in Fig. 8, and this would create an eddy  current in the $+\hat{y}$ direction generating entropy and rendering the process irreversible.
 Instead, if the normal state carriers (n carriers) are positive holes, electric and magnetic forces exactly cancel each other if the hole carriers
 are moving at the same speed $\dot{x}_0(t)$ as the phase boundary, as given by Eq. (29) and shown in Fig. 8. 

This then implies that the normal carriers of materials  that become superconducting in a reversible 
process and exhibit a Meissner
effect are necessarily $holes$.

                            \begin{figure}
 \resizebox{8.5cm}{!}{\includegraphics[width=6cm]{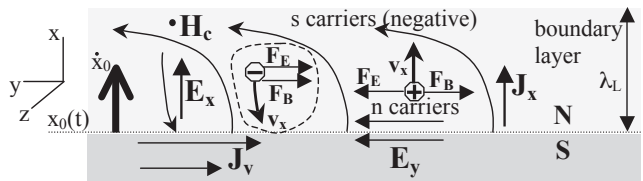}}
 \caption { Analysis of forces on the normal backflowing charge carriers (n carriers). If the carriers are electrons,
 electric and magnetic Lorentz forces $F_E$ and $F_B$ act in the same direction creating a current in the $\hat{y}$ direction.
 We argue that this does not occur, hence have circled such a carrier with a dashed line.
 Instead, if the normal carriers are holes, electric and magnetic forces act in opposite directions and
 exactly cancel each other, generating no current   in the $\hat{y}$ direction. We argue that this is the
 situation in real materials.}
 \label{figure1}
 \end{figure} 
           
      \section{resolution of the angular momentum puzzle}
      
      For several years    we have been pointing out that the Meissner effect raises a puzzling question
      concerning angular momentum conservation \cite{ang1,ang2}. Consider a superconducting
      cylinder with axis in the $\hat{z}$ direction  to which a magnetic field in the $+\hat{z}$ direction is applied.      Experiments show \cite{gyro}  that the body as a whole develops angular momentum in the  $-\hat{z}$
      direction, consistent with the fact that electrons in the Meissner current have angular momentum
      in the $+\hat{z}$ direction to generate a magnetic field in the $-\hat{z}$ direction that nullifies the
      field in the interior. For this situation the development of angular momentum for both the 
      electrons and the ions can be understood as arising from the force created on the 
      charges by the Faraday electric field generated by the changing magnetic field attempting to
      penetrate the superconductor \cite{ang1}.
      However, for the reverse situation where a metallic cylinder is cooled into the superconducting state
      in the presence of a magnetic field, the same Meissner current results, hence the same angular
      momentum has to be generated for both the electrons and the ions respectively. This has not been
      tested experimentally but is dictated by conservation of angular momentum. In this case however
      the motion of both the negative electrons in the Meissner current
       and the positive ions is in direction opposite to that dictated by the Faraday
      electric field, as shown in Fig. 9. We have called this
      the angular momentum puzzle.
      
      In Ref. \cite{ang2}, titled `the missing angular momentum of superconductors', we discussed this question
      and argued that it can be explained through the role of the spin-orbit interaction in the superconducting
      transition. While we still believe that the spin-orbit interaction plays a key role in superconductivity \cite{bohr},
      we don't believe that it is the explanation for the angular momentum puzzle. 
      
      In Ref.  \cite{dyn2}  we have proposed that the angular momentum puzzle is resolved 
      through transfer of momentum of the backflowing normal electrons to the lattice through 
      scattering by impurities. However, we don't believe that this is the solution to the
      angular momentum puzzle either.

                                              \begin{figure}
 \resizebox{5.5cm}{!}{\includegraphics[width=6cm]{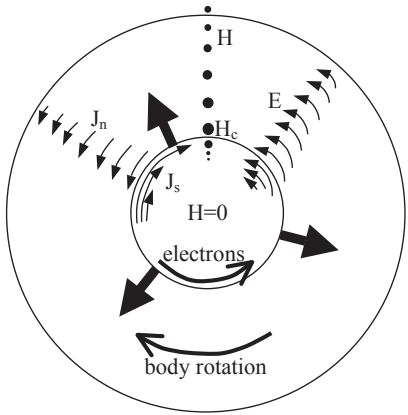}}
 \caption { Superconducting phase expanding from the center in a cylindrical geometry, with magnetic field
 pointing out of the paper.
 Electrons in the Meissner current flow in counterclockwise direction  and their angular momentum
 increases as the phase boundary moves out. Correspondingly, the body rotates in the clockwise direction
 with increasing angular velocity as the phase boundary moves out.
 }
 \label{figure1}
 \end{figure}

      Consider once more the process of backflow in the planar geometry, shown now in Fig. 10.
      In a sense, whether we talk about holes or electrons is semantics. In the process of a hole moving
      in the $+\hat{x}$ direction an electron necessarily has to be moving in the $-\hat{x}$ direction.
      The motion is exactly along the $\hat{x}$ direction because we have argued that the forces in the
      $\hat{y}$ direction are balanced for hole carriers.
      
      But the electric and magnetic forces point in the same direction for the negative electron. How is it
      possible that it moves purely in the $-\hat{x}$ direction?
      
      The answer is, of course, that there is another force acting on the electron.
      {\it The lattice exerts a force on the electrons when the charge carriers are holes}. 
      In order to balance the forces in the $-\hat{y}$ direction on the electrons, the lattice
      has to exert a force $F_L=F_E+F_B$ on the electron, pointing in the $+\hat{y}$ direction, as shown in Fig. 10.
      
                                              \begin{figure}
 \resizebox{8.5cm}{!}{\includegraphics[width=6cm]{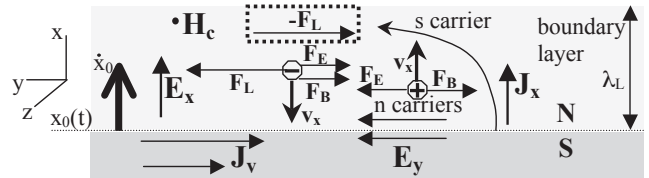}}
 \caption { Balance of forces. Holes propagate in the $+\hat{x}$ direction, which means electrons propagate
 in the $-\hat{x}$ direction. In order for this to happen, the lattice has to exert a force $F_L=F_E+F_B=2F_E$
 on the electrons in the $+\hat{y}$ direction, shown in the figure. This implies that the electrons are exerting a force $-F_L$ on the lattice in the $-\hat{y}$
 direction, shown in the figure in the dotted rectangle. In addition,
 the Faraday field $E_y$ exerts a force on the positive lattice that is  half as large as $F_L$
 and points in the $+\hat{y}$ direction (not shown in the figure).
 The net result is that a force $F_E$ in the $-\hat{y}$ direction acts on the lattice.
 }
 \label{figure1}
 \end{figure} 
      
      And, by Newton's 3rd law, the electrons in the `backflow' current exert a force
      $-F_L=-(F_B+F_E)$ pointing in the $-\hat{y}$ direction, also shown in Fig. 10.
      
      In addition, the electric field $E_y$ exerts directly a force $F_E$ on the positive ions in the crystal
      that points in the $+\hat{y}$ direction and is half the magnitude of $F_L$ (not shown in Fig. 9). The net result is,
      a net force $F_E$ is exerted on the lattice in direction $-\hat{y}$. This net force transfers to the lattice exactly the same
      momentum (in opposite direction)  that is acquired by the electrons becoming superconducting and
      carrying the Meissner current.
       The same reasoning explains the transfer of angular momentum to the body in the cylindrical geometry.
           
      This then explains how the lattice can acquire a momentum (or angular momentum) opposite to that
      dictated by Faraday's law when the Meissner current is generated, and resolves the `angular momentum puzzle'. The key to the solution is,
      {\it the carriers in the normal state have to be holes}.
      
      Note that this transfer of momentum to the lattice  is a non-dissipative process, different from the
      process that we had proposed earlier involving scattering by impurities \cite{dyn2}. It generates no entropy
      and the process remains reversible.
      
                                                                      \begin{figure}
 \resizebox{7.5cm}{!}{\includegraphics[width=6cm]{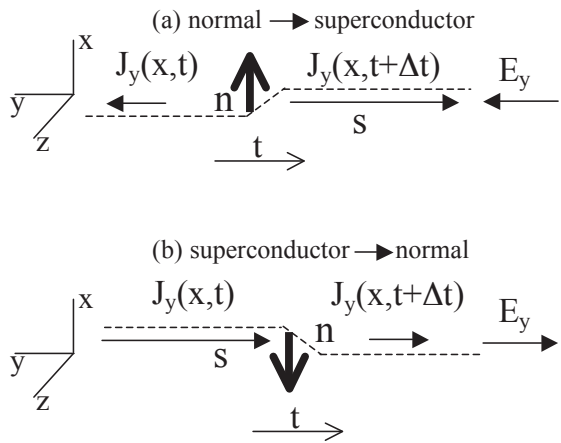}}
 \caption {(a) Phase boundary moves into the superconducting region, and (b) phase boundary moves into the normal region.
The induced electric fields $E_y$ point in opposite directions and so do the induced Foucault currents in the conventional description.
In the superconducting region the current is the same for both cases.
 . }
 \label{figure1}
 \end{figure}

            \section{experimental test}
            
            Consider the growth of the superconducting phase in the presence of a magnetic field in a planar geometry as shown in Fig. 12.
            We have argued that a  hole current $J_x$ flows in a boundary layer of thickness $\lambda_L$
             in front of the moving phase boundary, as shown in Fig. 12 (a). 
            Electric and magnetic forces on the normal hole carriers in the $\hat{y}$ direction are balanced, as was shown in Fig. 8. 
            This gives rise to a `Hall voltage' $V_H$ in the $\hat{y}$ direction, that can  be detected  by placing contacts on the sample along a line parallel to the  phase boundary that is approaching at speed $\dot{x}_0$.  If the distance
            between the contacts is $d$, the voltage measured will be $V_H=(H_c/c)\dot{x}_0 d$.
            
            For example, for $H_c=200G=60,000V/cm$, the boundary moving at speed $\dot{x}_0=0.1 cm/s$, and distance between
contacts  $d=1cm$, the measured voltage will be $V_H=0.6 \mu V$. The voltage will appear during the time interval $\Delta t$ where the boundary layer of thickness $\lambda_L$ is moving across the 
 $x$-position where the contacts are, $\Delta t=\lambda_L/\dot{x}_0$ . For $\lambda_L=500A$, $\Delta t=50 \mu s$.
 The polarity of the measured voltage will be as shown in Fig. 12, corresponding to a $positive$ Hall voltage
 originating in conduction by holes.
 In contrast, if there is no charge flow in the $\hat{x}$ direction as predicted by the conventional theory, the Faraday electric field
 $E_y$ will cause charge to accumulate at the lateral edges of the sample, as shown in Fig. 12 (b), so as to exactly compensate
 the Faraday field $E_y$ (positive (negative) charge to the left (right)), and the voltmeter will measure zero voltage throughout the process.
 
 For other geometries, similar differences in the expected results of such measurements predicted by our theory and the conventional theory may be expected.
            
                                               \begin{figure}
 \resizebox{8.5cm}{!}{\includegraphics[width=6cm]{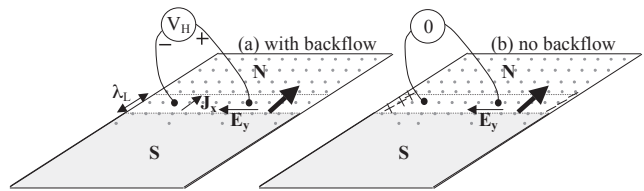}}
 \caption { In the presence of backflow, a voltmeter connected to the sample as shown in (a) will detect a Hall voltage $V_H$ proportional
 to the speed of motion of the phase boundary, during the time interval where the boundary layer of thickness $\lambda_L$ in front of the
 phase boundary moves across the region of the contacts. In the absence of backflow (b), no voltage will be measured.
 }
 \label{figure1}
 \end{figure}
      \section{summary and discussion}

In previous work \cite{dyn1,dyn2} we have argued that to understand the dynamics of the Meissner effect it is necessary to assume that
there is motion of charge in direction perpendicular to the normal-superconductor phase boundary, which is not expected within the
conventional BCS-London theory of superconductivity. In this paper we have pointed out that such motion of charge also explains
why the normal-superconductor transition is experimentally found to be reversible to high accuracy: the magnetic Lorentz force on 
normal charge
carriers moving perpendicular to the phase boundary in a boundary layer cancels the tangential force due the the Faraday electric field that necessarily
arises when the phase boundary moves, thus supressing the generation of Joule heat by Foucault currents flowing parallel to the
phase boundary.  The physics discussed in this paper also explains how the Meissner current disappears in the process of the superconductor becoming
 normal without being dissipated as Joule heat.

                                                          \begin{figure}
 \resizebox{7.5cm}{!}{\includegraphics[width=6cm]{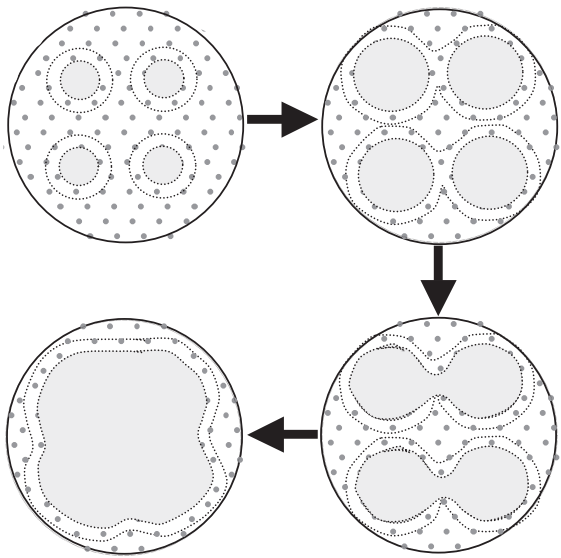}}
 \caption {Transition to the superconducting state (schematic) in the presence of a magnetic field (grey dots). 
 Superconducting domains (grey areas) start at random places and grow outward, surrounded by a boundary layer (dotted lines)
 where current normal to the phase boundary exists and 
 within which no tangential currents exist. The domains expand and merge until they occupy the entire sample. }
 \label{figure1}
 \end{figure}
 
It should be pointed out that some degree of irreversibility will always exists if there are regions of the sample that are not part of boundary
layers, since in those regions eddy currents will flow. This will  certainly be the case if for example a single domain grows from the center
in a cylindrical sample as shown in Fig. 9. However if the growth of the superconducting phase occurs by formation of many separate domains that
grow and subsequently merge, as shown schematically in Fig. 13, the fraction of the sample occupied by boundary layers correspondingly 
grows and Joule heating is reduced because no eddy currents flow within the boundary layers. 
Thus, one can envision a scenario where a sufficient number of domains start growing simultaneously so that any point in the sample is
either in the superconducting state or in a boundary layer at any time, in which case entropy will not be generated and the transition will be
completely reversible and take a finite amount of time. This is $qualitatively$  different from the conventional description where
the transition $cannot$ occur without entropy production.

Furthermore we  showed that this suppression of eddy currents that would otherwise render the process irreversible will only occur if the carriers in the normal state
are $holes$. Notably, this agrees with the prediction of the theory of hole superconductivity \cite{holesc}
that superconductivity can only occur if the normal state current carriers are holes.

 Furthermore, we showed that this physics also explains the angular momentum puzzle pointed out in previous work \cite{ang1,ang2}.
 The flow of normal carriers in direction perpendicular to the phase boundary imparts momentum (or angular momentum) to the solid as a whole that exactly
 compensates the momentum (or angular momentum) gained by the carriers that become superconducting and carry the Meissner current, or lost by the superconducting carriers in the Meissner current that become
 normal. For this to
 happen it is also a necessary requirement that the carriers in the normal state are holes. This is because momentum transfer to the
 solid does not occur through scattering processes as we had earlier hypothesized \cite{ang2,dyn1}, but simply due to the 
 fact that electrons of negative effective mass (holes) necessarily transfer momentum to the lattice when current flows \cite{holeelec2,kittel}.
 
 An essential aspect of the physics discussed here is that the wavefunction (`orbit') and the charge of superconducting electrons
 close to the normal-superconductor phase boundary penetrates into the normal region. 
 It is likely that this physics is related to the well known proximity effect, where signs of weaks superconductivity are
 observed in junctions of superconducting and normal metals \cite{prox}. 
 In the conventional explanations of that effect the superconducting order parameter is asumed to extend into
 the normal region \cite{prox2}  but this is not expected to be also associated with transfer of negative charge from the
 superconducting to the normal region, as in the physics described in this paper. 
 In future work we will explore the applicability of the physics discussed here to describe proximity effect
 phenomena.

 As discussed in this paper, the way in which a metal expels a magnetic field from its interior in the process
 of becoming superconducting, while conserving angular momentum and  overcoming Faraday's law,  is far from trivial. We had pointed out in earlier work that 
 expulsion of magnetic field necessarily has to involve   a radial outward 
 motion of charge \cite{outward}. However, as discussed in ref. \cite{dyn2},  if it involved only outward motion of charge it would require the entire 
 mobile charge in the metal to move out from the interior across the surface of the sample carrying the magnetic field lines with it, which obviously 
 does not happen.
 Instead, the superconductor achieves this feat in a rather elaborate  way, analogous to the mechanism of a ratchet wrench: the unrestricted motion
 corresponds to the outflow of  electrons becoming superconducting over a distance $\lambda_L$, followed (although it happens concurrently) by the backflow of normal electrons that because of their antibonding character
 (negative effective mass)
 apply a force (or torque) to the entire body, and this combined  flow and counterflow repeated over and over carries the magnetic field lines gradually out of the
 body over a macroscopic distance while  conserving total momentum and angular momentum. In a sense it is as if all the mobile carriers flow out of the body as electrons and flow in again as antibonding electrons, or holes.
 It is not surprising that this process can take an extended period of time. It is also not surprising that the end result of this large
 amount of charge flowing out as light electrons and backflowing as heavy holes leaves as end result a small charge imbalance where
  some  excess negative charge remains within the London penetration depth of the surface of the superconductor and some excess positive
 charge in the interior as predicted by the theory of hole superconductivity \cite{chargeexp}.

 Instead, the conventional BCS-London theory of superconductivity explains the reversibility of the normal-superconductor transition in the presence of induced Faraday fields, the dynamics of the 
 generation of the Meissner current in apparent violation of 
 Faraday's law, the transfer of momentum from the charge carriers to the  lattice required to conserve momentum and angular momentum, and the disappearance of the Meissner current without
 irreversible heat loss, in a much simpler way: by simply postulating that it happens and that it needs no further explanation \cite{bcs50}.

 One way to test the physics discussed in this paper would be to repeat the experiments measuring the latent heat in the superconducting
 transition in a field \cite{keesom,keesom2} with very high purity samples \cite{tinres}, and verify that reversibility is satisfied
 to high accuracy for experiments extending over a period of time where the conventional theory would predict a much higher
 degree of irreversibility, as discussed in Sect. V. Another way would be to detect Hall voltages as the phase boundary advances that would not be
 expected within the conventional theory, as discussed in Sect.    X.

\end{document}